\documentclass[aps,nofootinbib,superscriptaddress,floatfix]{revtex4} 
\usepackage{amsmath}
\usepackage{graphicx,epsfig,wrapfig}
\usepackage{amsthm}
\usepackage{makeidx}
\usepackage{amsfonts}
\usepackage{amssymb}
\usepackage{amsmath}
\usepackage{natbib}
\usepackage{dsfont}
\usepackage[capitalize]{cleveref}
\usepackage{mathrsfs}
\usepackage{slashed}
\usepackage{epstopdf}
\usepackage{bm}
\usepackage{color}
\usepackage[normalem]{ulem}
\usepackage{simplewick}

\usepackage{multirow}
\usepackage{braket}\newcommand{\la}{\langle}
\renewcommand\sout{\bgroup \color[rgb]{0.55,0.00,0.99} \ULdepth=-.5ex \ULset}

\begin{document}

\newcommand{\ud}{\text{d}}
\newcommand{\mb}[1]{\boldsymbol{#1}}
\newcommand{\mc}[1]{\mathcal{#1}}
\newcommand{\non}{\nonumber}
\newcommand{\pep}[1]{\mathbf{#1}_{\perp}}
\newcommand{\pepgr}[1]{\bm{#1}_{\perp}}
\newcommand{\gdir}[1]{\gamma^{#1}}
\newcommand{\xk}{(x,\mathbf{k}_{\perp})}
\newcommand{\xkq}{(x,\mathbf{k}_{\perp}^{2})}
\newcommand{\pe}{(\mb{p}_{e})}
\newcommand{\xks}{(x,\pep{k};S)}
\newcommand{\pv}{\mathrm{P.V.}}
\newcommand{\xkl}{(x,\pep{k};\Lambda)}
\newcommand{\dEF}[1]{\nabla^{\text{EF}}_j\left(#1\right)}

\newcommand{\ra}{\rangle}

\newcommand{\di}{\mathrm{d}}
\newcommand{\ta}{\left(}
\newcommand{\qa}{\left[}
\newcommand{\ga}{\left\{}
\newcommand{\tc}{\right)}
\newcommand{\qc}{\right]}
\newcommand{\gc}{\right\}}
\newcommand{\ma}{\left|}
\newcommand{\mnu}[1]{{#1}^{\mu \nu}}
\newcommand{\mnd}[1]{{#1}_{\mu \nu}}
\newcommand{\prodotto}[2]{\ta {#1}\cdot {#2} \tc }
\newcommand{\dq}[1]{\delta^{(4)}\left({#1}\right)}
\newcommand{\dt}[1]{\delta^{(3)} \ta {#1} \tc}
\newcommand{\du}[1]{\delta \ta {#1} \tc}
\newcommand{\vet}[1]{\bm{{#1}}}
\newcommand{\vetp}[1]{\bm{{#1}_{\perp}}}
\newcommand{\pp}{\vet{p}}
\newcommand{\qq}{\vet{q}}
\newcommand{\radice}[1]{\sqrt{{#1}}}
\newcommand{\tr}[1]{\mbox{Tr}\qa {#1} \qc}

\newcommand{\anti}[2]{\overline{{#1}}\ta {#2}\tc }
\newcommand{\antif}[2]{\overline{{#1}}^{(f)}\ta {#2}\tc }
\newcommand{\antim}[2]{\overline{{#1}}_-^{(f)}\ta {#2}\tc }
\newcommand{\antip}[2]{\overline{{#1}}_+^{(f)}\ta {#2}\tc }

\title{The twist-three distribution $e^q(x,k_\perp)$ in a light-front model}

\author{Barbara Pasquini}
\email{barbara.pasquini@pv.infn.it}
\affiliation{Dipartimento di Fisica, Universit\`a degli Studi di Pavia, I-27100 Pavia, Italy}
\affiliation{Istituto Nazionale di Fisica Nucleare, Sezione di
  Pavia,  I-27100 Pavia, Italy}

\author{Simone Rodini}
\email{simone.rodini01@ateneopv.it}
\affiliation{Dipartimento di Fisica, Universit\`a degli Studi di Pavia, I-27100 Pavia, Italy}
\affiliation{Istituto Nazionale di Fisica Nucleare, Sezione di
  Pavia,  I-27100 Pavia, Italy}

\date{\today}
\bibliographystyle{myrevtex}
\allowdisplaybreaks[2]

\begin{abstract}
We discuss
the twist-three, unpolarized, chiral-odd, transverse momentum dependent parton distribution (TMD) $e^q(x,k_\perp)$ within a light-front model.
We review a model-independent decomposition of this TMD, which follows from the QCD equations of motion and is
given  in terms of a leading-twist mass term, a pure interaction-dependent contribution, and singular terms.
The leading-twist and pure twist-three terms are represented in terms of overlap of light-front wave functions (LFWFs), taking into account the Fock states with three valence quark ($3q$)  and  three-quark plus one gluon ($3q+g$).
The $3q$ and $3q+g$ LFWFs with total orbital angular momentum zero are modeled using a parametrization derived from the conformal expansion of the proton distribution amplitudes, 
with parameters fitted to reproduce
available phenomenological information on the unpolarized leading-twist quark and gluon collinear  parton distributions.
Numerical predictions for both the quark TMD $e^q(x,k_\perp)$ and the collinear parton distribution $e^q(x)$ are presented, discussing the role of the quark-gluon correlations in the proton.
\end{abstract}

\maketitle

\section{Introduction}

Higher-twist  parton distributions, especially when the dependence  on the parton transverse momenta is taken into account, 
give access to a wealth of information about the nucleon parton structure~\cite{Jaffe:1983hp,Mulders:1995dh,Goeke:2005hb}.
They describe multiparton correlations inside the nucleon, corresponding to the interference between scattering from a coherent quark-gluon pair and from a single quark~\cite{Jaffe:1991ra,Jaffe:1991kp,Efremov:2002qh,Burkardt:2008ps}.
As such, they help understanding the quark-gluon dynamics inside the hadrons, and go beyond the probabilistic  interpretation that applies to the leading-twist parton distribution functions.

Twist-three parton distributions functions (PDFs) and transverse momentum dependent parton distributions (TMDs) contribute to various observables in inclusive and semi-inclusive deep inelastic scattering (SIDIS), respectively.
Although suppressed with respect to twist-two observables, twist-three structure functions are
not small in the kinematics of fixed target experiments. 
One of the priority tasks of the future experimental program at JLab12 is the measurement
of different higher-twist spin-azimuthal asymmetries in SIDIS~\cite{E12017,E12-06-112,E12-06-112B}.
A future electron ion collider  would extend such experimental investigation by accessing different kinematical regions~\cite{Boer:2011fh,Accardi:2012qut}.

In this context, model studies have been shown to have important impact for the understanding of TMDs and the theoretical interpretation of related observables (see, e.g., Ref.~\cite{Metz:2004ya}).
Higher-twist quark PDFs and TMDs can in general be decomposed into contributions from leading-twist mass  terms, singular terms and pure interaction-dependent (``tilde") terms.
This decomposition  is obtained through  the QCD equations of motion (EOM) and allows one to single  the tilde term out as  the contribution of quark-gluon correlation functions.
Neglecting the tilde and mass terms is  referred to as Wandzura-Wilczek approximation~\cite{Wandzura:1977qf}. This approximation has been often used
as starting point to simplify the description of twist-three SIDIS observables~\cite{Efremov:2002ut,Efremov:2002td,DeSanctis:2000fh}, and  showed to be an useful numerical approximation~\cite{Avakian:2007mv}. However, 
there is no real experimental evidence of its validity, and
it misses one of the main motivation to study sub-leading twist, i.e. the 
non-perturbative physics of quark-gluon correlations. 
 
 Twist-three TMDs have been calculated in various models:  the MIT bag model~\cite{Jaffe:1991ra,Signal:1996ct,Avakian:2010br,Lorce:2014hxa}, 
diquark spectator models~\cite{Jakob:1997wg,Lu:2012gu,Mao:2013waa,Mao:2014aoa},
 instanton models of QCD vacuum~\cite{Balla:1997hf,Dressler:1999hc}, chiral quark soliton models~\cite{Schweitzer:2003uy,Wakamatsu:2007nc,Wakamatsu:2003uu,Ohnishi:2003mf,Cebulla:2007ej}  and perturbative light-front
Hamiltonian approaches with a quark target~\cite{Burkardt:2001iy,Kundu:2001pk,Mukherjee:2009uy,Accardi:2009au}.
Although quark models do not have explicit gluon degrees of freedom, they describe interacting quarks
and can generate non-vanishing tilde terms~\cite{Lorce:2014hxa,Lorce:2016ugb}.

In this work we want to make a step forward with respect to quark-model calculations and take into account explicitly 
the contribution from intrinsic gluon degrees of freedom.
To this aim, we use the language of light-front wave functions (LFWFs), that provide a convenient framework  for modelling parton distribution functions~\cite{Diehl:1998kh,Lorce:2011dv,Burkardt:2015qoa}.
We focus on the twist-three, unpolarized, chiral-odd quark TMD $e^q(x,k_\perp)$~\cite{Efremov:2002qh},
which is constructed as  overlap  integrals between LFWFs with the minimum (valence) and next-to-minimum
(one extra gluon) parton content.
The LFWFs are modeled in the same spirit of Ref.~\cite{Braun:2011aw}, where the calculation was restricted to the leading-twist PDFs and to the twist-three polarized structure function $g^q_2(x)$.
In particular, we consider only the Fock states with zero partons' orbital angular momentum, which are expected to be the dominant contribution for unpolarized distribution functions. 
These LFWF components for the $3q$ and $3q+g$ Fock states are related, in the light-front limit (zero transverse separation), to the nucleon twist-three and twist-four distribution amplitudes (DAs), respectively. 
We then use the lattice and  QCD sum rule results for the proton DAs as guideline to parametrize the dependence of the LFWFs on the longitudinal momenta of the partons.
 For the dependence on the partons' transverse momentum $k_\perp$ we adopt a  Gaussian form, modified according to the Brodsky-Huang-Lepage prescription~\cite{BHL}
 to take into account a non-vanishing  mass of the partons.
The mass of the partons along with the other parameters modelling the proton DAs are then fitted to reproduce  the results for the quark and gluon unpolarized twist-two PDFs
from available phenomenological parametrizations. Having specified the LFWFs, we calculate the $e^q(x,k_\perp)$ TMD and $e^q(x)$ PDF, discussing the role of the twist-two and the genuine twist-three contributions, and compare our  predictions with available phenomenological information.

The work is organized as follows: in Sec. \ref{sec:1} we introduce the definitions of the unpolarized twist-two TMDs of quark and gluon  $f^{q/g}_1(x,k_\perp)$ and the twist-three quark TMD $e^q(x,k_\perp)$, and we review the general decomposition of $e^q(x,k_\perp)$ derived from the QCD EOM. In Sec. \ref{sec:2}, we present the  Fock-state expansion of the proton state, and describe a model-independent representation of the LFWFs for  the $3q$ and $3q+g$ components with zero partons' orbital angular momentum, as derived originally in Refs.~\cite{Ji:2003yj,Braun:2011aw}.
Then, we construct the corresponding LFWF overlap representation for the pure twist-three contribution to $e^q(x,k_\perp)$ and introduce our parametrization for the LFWFs, in terms of proton DAs. In Sec. \ref{sec:3}, we fix the model parameters of the LFWFs by fitting the quark and gluon  PDF $f_1(x)$ to the  MMHT2014 parametrization~\cite{Harland-Lang:2014zoa} at the scale of 1 GeV$^2$, and discuss our model predictions for both the TMD $e^q(x,k_\perp)$ and the PDF $e^q(x)$. We summarize the work in Sec.~\ref{sec:4}. Technical details about the derivation of the model-independent decomposition of $e^q$ are given in App.~\ref{appendix-1}, and the expression of  $e^q(x,k_\perp)$ in terms of our model LFWFs can be found in App.~\ref{appendix-2}.

\section{Definition of unpolarized distributions and decomposition of $e^q(x,k_\perp$)}
\label{sec:1}
Quark  TMDs are defined from the twist expansion of the following quark-quark correlators (see, e.g., Refs. \cite{Jaffe:1996zw,Goeke:2005hb}).
\begin{align}
\Phi^q(x, &k_\perp,\,S)  = \int \frac{dz^-d\vetp{z}}{(2\pi)^3} e^{iz^-k^+ - i\vetp{z}\cdot \vetp{k}} \braket{P,S|\overline\psi(0)\mathcal{W}(0;z)\psi(z)|P,S}|_{z^+ = 0}, \label{quark-correlator}
\end{align}
with $k^+=xP^+$~\footnote{We use light-front coordinates,  with   $v^\pm=\tfrac{1}{\sqrt{2}}(v^0\pm v^3)$ and $\vetp{v}=(v^1, v^2)$ for a generic four-vector $v$.}.
The target state is characterized by its four-momentum
$P$ and the covariant spin vector $S$. In Eq.~\eqref{quark-correlator}, $\psi$ is the quark field and $\mathcal{W}$ is an appropriate Wilson line, that connects the bilocal quark operators and ensures gauge invariance~\cite{Bomhof:2004aw}. It is defined as:
\begin{align}
\mathcal{W}(0,z)&=[0^+,0^-,\bold{0}_\perp;0^+,\infty^-,\bold{0}_\perp]
\times[0^+,\infty^-,\bold{0}_\perp;z^+,\infty^-,\boldsymbol{\infty}_\perp]\nonumber\\
&
\times[z^+,\infty^-,\boldsymbol{\infty}_\perp;z^+,\infty^-,\bold{z}_\perp]
\times[z^+,\infty^-,\bold{z}_\perp;z^+,z^-,\bold{z}_\perp]
\end{align}
where
$[a^+,a^-,\bold{a}_\perp;b^+,b^-,\bold{b}_\perp]$ denotes a gauge link connecting the points $a^\mu=(a^+,a^-,\bold{a}_\perp)$ and $b^\mu=(b^+,b^-,\bold{b}_\perp)$ along a straight line.

At twist-three level, we find three unpolarized T-even quark TMDs, i.e. the twist-two TMD $f^q_1(x,k_\perp)$, and the twist-three TMDs $e^q(x,k_\perp)$ and $f^\perp(x,k_\perp)$.
In the following, we restrict ourselves to discussing the TMDs $f_1$ and $e$,  which do not involve partons' orbital angular momentum transfer between the initial and final states.
They are defined in terms of the quark-quark correlator as
  \begin{align}
 &f_1^q(x,k_\perp)=  \Phi^{[\gamma^+]}=\frac 12\int\di k^- \,{\rm tr}[\Phi^q\gamma^+]
   = 
   \int \frac{\di z^-\di^2z_\perp}{2(2\pi)^3}  
   e^{i k \cdot z} \, \la P  |  \overline{\psi}(0)
   \mathcal{W}(0;z)\gamma^+
   \psi(z)  |  P\ra|_{z^+=0} ,\label{Eq:correlator-TMDs1}
   \end{align}
   \begin{align}
 &\frac{M}{P^+}\,e^q(x,k_\perp)   = \Phi^{[\mathds{1}]}=\frac 12\int\di k^- \,{\rm tr}[\Phi^q\mathds{1}]
   = 
   \int \frac{\di z^-\di^2z_\perp}{2(2\pi)^3} 
   e^{i k \cdot z} \, \la P  |  \overline{\psi}(0)
   \mathcal{W}(0;z)\mathds{1}
   \psi(z)  |  P\ra|_{z^+=0},  \label{Eq:correlator-TMDs2}
\end{align}
where $\langle P|\dots|P\rangle $ denotes the target spin-averaged matrix element and $M$ is the nucleon mass.
The  quark collinear PDFs $f^q_1(x)$ and $e^q(x)$ are obtained 
by integrating the corresponding TMDs over the transverse momentum $\vetp{k}$.

For later convenience, we also introduce  the gluon unpolarized twist-two PDF $f_1^g(x)$, defined in terms of  the gluon correlator  $\Phi^{g,ji}(x,  \vet{k}_{\perp})$ as
\begin{align}
& f_1^g(x,k_\perp) = -g_{ij} \Phi^{g,ji}(x,k_\perp)
 = \sum_{a=1}^8\sum_{i=1}^{2} \frac{1}{xP^+} \int \frac{dz^-d\bm{z}_{\perp}}{(2\pi)^3} e^{ik\cdot z}
\braket{P,S| \mathcal{W}(z;0)F^{+i}_a\ta 0\tc \mathcal{W}(0;z) F^{+i}_a(z)|P,S}\Big|_{z^+ = 0}, 
\label{f1gDef}
\end{align} 
where  $F^{\mu\nu}_a$ is the gluon field strength tensor and $a$ is the color index.

We now focus our attention  on the quark TMD $e^q(x,k_\perp)$,  and review its general  decomposition  in terms of leading-twist  quark-mass terms, singular terms
and pure interaction-dependent contributions.
The bilocal quark operator entering the definition  in Eq.~\eqref{Eq:correlator-TMDs2} can be rewritten as 
\begin{align}
\mathcal{O}\ta 0; z^-,\vetp{z}\tc &= \overline{\psi}(0)\mathcal{W}(0;z)
   \psi(z)|_{z^+=0}   = \overline\psi_+(0)\mathcal{W}(0;z)\psi_-(z)|_{z^+=0}
   + \overline\psi_-(0)\mathcal{W}(0;z)\psi_+( z)|_{z^+=0} ,
\label{defO}
\end{align}
where we introduced the projection of the quark fields into the light-cone `good' and `bad' components, i.e.,  $\psi_+=\mathcal{P}^+\psi = \psi_+$ and  $\psi_-=\mathcal{P}^-\psi$, respectively, with $\mathcal{P}^{\pm}= \frac{1}{2}\gamma^{\mp}\gamma^{\pm}$.
The bad component $\psi_-$ is a constrained field, as follows from the QCD EOM: 
\begin{equation}
iD^+\psi_-(z) = \frac{\gamma^+}{2}\ta i\vetp{\gamma}\cdot \vetp{D} + m \tc \psi_+(z),\label{EOM}
\end{equation}
where the covariant derivative is defined as
\begin{equation}
D^{\mu} = \partial^{\mu} - ig_s A^{\mu},\label{covariant}
\end{equation}
with $g_s$ the strong coupling constant.
If we assume that the plus component $k^+$ of the quark's momentum is strictly positive, 
we can invert
the EOM \eqref{EOM} in a straightforward way using the Fourier expansion of the fields. Instead, problems arise when we include the contribution from zero modes corresponding to $k^+=0$. In this case,
there appear  singularities in the bad components of the field, and one needs   a regularization prescription~\cite{Zhang:1993is,Zhang:1993dd}.
To this aim, we  follow the procedure outlined in Refs.~\cite{Efremov:2002qh,Kodaira:1998jn} and  use  the EOM \eqref{EOM}  to derive  the following operator identity 
\begin{align}
\mathcal{W}(0;z)\,\psi_-(z))|_{z^+=0}
&= \mathcal{W}_1(0^-,\bold{0}_\perp;0^-,\vetp{z})\psi_-(0^+,0^-,\vetp{z}) \notag \\
& - i \int_{0^-}^{z^-} d\zeta^- \mathcal{W}_1(0^-,\bold{0}_\perp;\zeta^-,\vetp{z}) 
\frac{\gamma^+}{2}\ta i\vetp{\gamma}\cdot \vetp{D} + m \tc \psi_+(0^+,\zeta^-,\vetp{z}),
\label{PSI_Int}
\end{align}
where we defined $\mathcal{W}_1(a^-,\bold{a}_\perp;b^-,\bold{b}_\perp)\equiv\mathcal{W}(0^+,a^-,\bold{a}_\perp;0^+,b^-,\bold{b}_\perp)$.
Using the expression in Eq.~\eqref{PSI_Int} for the bad component, Eq.~\eqref{defO} can be rewritten as
\begin{align}
&\mathcal{O}\ta 0;  z^-,\vetp{z}\tc = \mathcal{O}_s+\mathcal{O}_m+\mathcal{O}_{\text{tw3}},
\end{align}
with
\begin{align}
\mathcal{O}_s&=\overline{\psi}(0)\mathcal{W}_1(0^-,\bold{0}_\perp;0^-,\vetp{z})\psi(0^+,0^-,\vetp{z}), \label{Os} \\
\mathcal{O}_m&= - i m\int_{0}^{z^-} d\zeta^- \overline \psi_+(0) \mathcal{W}_1(0^-,\bold{0}_\perp;\zeta^-,\vetp{z})\gamma^+\psi_+(0^+,\zeta^-,\vetp{z}), \label{Om} \\
\mathcal{O}_{\text{tw3}}&= - \frac{i}{2}  \int_{0}^{z^-} d\zeta^-  \overline \psi_+(0)\sigma^{j+} \Big[ \mathcal{W}_1(0^-,\bold{0}_\perp;\zeta^-,\vetp{z}) \overset{\rightarrow}{D}_{\perp,j}(\zeta^-,\vetp{z})
+ \overset{\leftarrow}{D}^\dagger_{\perp,j}(0)\mathcal{W}_1(0^-,\bold{0}_\perp;\zeta^-,\vetp{z})\Big]  \psi_+(0^+,\zeta^-,\vetp{z}). \label{Oa}
\end{align} 
where the index $j=1,2$ labels the transverse component.

The bad components $\psi_-$ contributes only to $\mathcal{O}_s$ in Eq.~\eqref{Os}. This operator, when inserted in the matrix element of Eq.~\eqref{Eq:correlator-TMDs2}
and integrated over $z^-$, gives a  singular contribution proportional 
 to $\delta(x)$:
 \begin{align}
&e^q_s\ta x,k_\perp\tc = \frac{\delta(x)}{2M}\int \frac{d\vetp{z}}{(2\pi)^2} e^{- i\vetp{z}\cdot\vetp{k}} 
\braket{P|
\overline{\psi}(0)\mathcal{W}_1(0^-,\bold{0}_\perp;0^-,\vetp{z})\psi(0^+,0^-,\vetp{z})
%
|P}.
\label{deltaTerm}
\end{align}
This contribution is well known for the PDF $e^q(x)$, being related to the pion-nucleon-sigma term (see, e.g., Ref.~\cite{Efremov:2002qh}). 

The contribution to $e^q$ from the operator $\mathcal{O}_m$ can be worked out using the Fourier expansion of the matrix element
of the operator~\eqref{Om}  and the definition of $f_1^q$  in~\eqref{Eq:correlator-TMDs1}, as outlined in App.~\ref{appendix-1}, with the result
\begin{align}
e^q_m&=  \frac{m}{Mx}f_1^q(x,k_\perp) - \frac{m}{M}\delta(x) \int_{-1}^1 dy \frac{f_1^q(y,k_\perp)}{y},
\label{MassTerm}
\end{align}
where the singular term   is a natural consequence of the  divergences associated with the zero modes ($x = 0$).

Limiting ourselves to the T-even sector and to the target-spin averaged matrix element,   the contribution from the operator $\mathcal{O}_{\text{tw3}}$ can be rewritten as
\begin{align}
e_{tw3}  &= -\frac{P^+}{M} \frac{g_s}{2}\int \frac{dz^-d\vetp{z}}{2(2\pi)^3}e^{ik^+z^--\bold{k}_\perp\cdot\bold{z}_\perp}\notag\\
&\times\Bigg(
  \int_{0^-}^{z^-} d\zeta^-\int_{\infty^-}^{\zeta^-} d\eta^-
   \braket{P|\overline\psi(0)\mathcal{W}_1(0^-,\bold{0}_\perp;\eta^-,\vetp{z})G_{\ j}^{+}(0^+,\eta^-,\vetp{z})\sigma^{j+}\mathcal{W}_1(\eta^-,\vetp{z};\zeta^-,\vetp{z})\psi(0^+,\zeta^-,\vetp{z})|P} \notag \\
   & +  \int_{0^-}^{z^-} d\zeta^-\int^{\infty^-}_{0^-} d\eta^- 
    \braket{P|\overline\psi(0)\mathcal{W}_1(0^-,\bold{0}_\perp;\eta^-,\vetp{0})G_{\ j}^{+}(0^+,\eta^-,\vetp{0})\sigma^{j+}\mathcal{W}_1(\eta^-,\vetp{0};\zeta^-,\vetp{z})\psi(0^+,\zeta^-,\vetp{z})|P}\Bigg),
    \label{FinalE}
\end{align}
where $G^{\mu\nu}$ is the gluon field strength tensor.
Using the results in App.~\ref{appendix-1}, Eq.~\eqref{FinalE} can be  recast in the form
\begin{align}
 &e^q_{\text{tw3}}(x,k_\perp)=\tilde e^q (x,k_\perp) -  \delta(x) \int_{-1}^{1}dy \,\tilde e^q(y,k_\perp),
\label{GluonTerm}
\end{align}
where\begin{equation}
\tilde e^q(x,k_\perp) = -\frac{i}{Mx}\Phi_{A,j}^{[\sigma^{j+}]}(x,k_\perp)\label{Etilde:def}
\end{equation}
is a pure twist-three contribution defined in terms of 
the quark-gluon-quark correlation function~\cite{Boer:2003cm}
\begin{align}
 \Phi_{A,j}^{[\sigma^{j+}]}(x,k_\perp)&=\frac{1}{2}\mathrm{Tr}\left[\Phi_{A,j}(x,k_\perp)\sigma^{j+}\right]   =\frac{g_s}{2} \int \frac{dz^-d\vetp{z}}{2(2\pi)^3}e^{ik^+z^--\bold{k}_\perp\cdot\bold{z}_\perp}\notag\\
&\times  \Bigg(
\int_{\infty^-}^{\zeta^-} d\eta^- \braket{P|\overline\psi(0)\mathcal{W}_1(0^-,\bold{0}_\perp;\eta^-,\vetp{z})G_{\ j}^{+}(0^+,\eta^-,\vetp{z})\sigma^{j+}\mathcal{W}_1(\eta^-,\vetp{z};\zeta^-,\vetp{z})\psi(0^+,\zeta^-,\vetp{z})|P} \notag \\
& +  \int^{\infty^-}_{0^-} d\eta^- \braket{P|\overline\psi(0)\mathcal{W}_1(0^-,\bold{0}_\perp;\eta^-,\vetp{0})G_{\ j}^{+}(0^+,\eta^-,\vetp{0})\sigma^{j+}\mathcal{W}_1(\eta^-,\vetp{0};\zeta^-,\vetp{z} )\psi(0^+,\zeta^-,\vetp{z})|P}\Bigg).
  \label{PhiADef} 
\end{align}

Collecting the results in Eqs.~\eqref{deltaTerm}-\eqref{GluonTerm}, we  end up with the following decomposition:
\begin{align}
e^q(x,k_\perp) & = e^q_s\ta x, k_\perp\tc + \tilde{e}^q\ta x,k_\perp\tc  + \frac{m}{xM}f^q_1\ta x,k_\perp\tc 
-  \delta(x) \int_{-1}^{1}dy \,\left(\frac{m}{My}f^q_1(y,k_\perp)+\tilde e^q(y,k_\perp)\right).
\label{genDecomp}
\end{align}
The singular term beyond the contribution of $e_s$ is usually not discussed in literature
and, to the best of our knowledge, has never been written explicitly in this form.

The decomposition~\eqref{genDecomp} is  independent on the choice of the gauge.
In the light-cone gauge $A^+ = 0$, with suitable boundary conditions at light-cone infinity for the transverse components of the gauge field, the gauge links
in the correlators can be ignored and $\Phi_{A,j}$ in Eq.~\eqref{PhiADef} is replaced by the following correlator~\cite{Boer:2003cm,Bacchetta:2006tn}
\begin{align}
\tilde{\Phi}_{A,j}^{[\sigma^{j+}]}(x,k_\perp)&=\frac{1}{2}\mathrm{Tr}\left[\tilde{\Phi}_{A,j}(x,k_\perp)\sigma^{j+}\right] =\frac{g_s}{2} \int \frac{dz^-d\vetp{z}}{2(2\pi)^3}e^{ik\cdot z}
    \braket{P,S|\overline{\psi}(0)\left[A_{\perp,j}(z)-A_{\perp,j}(0)\right]\sigma^{j+}\psi(z)|P,S}_{z^+ = 0}.
    \label{PhiADefLC} 
\end{align}

In the framework of light-front quantization in the $A^+=0$ gauge,
the effects of the final-state interactions associated with the gauge link can be reabsorbed in the LFWFs of the target, which
acquire an imaginary phase~\cite{Brodsky:2010vs}. 
As we are interested to T-even TMDs, these effects can be ignored, and 
in the following we will work only with real LFWFs.

Integrating Eq.~\eqref{genDecomp} over $\vetp{k}$, one obtains the corresponding decomposition
for the PDF $e^q(x)$
\begin{align}
e^q(x) & = e_s\ta x\tc + \tilde{e}^q\ta x\tc  + \frac{m}{xM}f^q_1\ta x\tc -  \delta(x) \int_{-1}^{1}dy \,\left(\frac{m}{My}f^q_1(y)+\tilde e^q(y)\right),
\label{genDecomp-PDF}
\end{align}
from which one can easily infer well-known relations for the  first Mellin moments of $e^q(x)$~\cite{Efremov:2002qh}.
 

\section{Light-front Fock-state expansion}
\label{sec:2}
In this section, we derive the LFWF overlap representation for the TMD $e^q(x,k_\perp)$.
Since we will work under the assumption $k^+>0$, in the following we will not consider the singular terms in Eqs.~\eqref{genDecomp} and \eqref{genDecomp-PDF}.

In the framework of light-front quantization in the $A^+=0$ gauge, and
restricting ourselves to the contributions from the $3q$ and $3q+g$ Fock-states, the light-front Fock-state expansion of a proton state with momentum $P$ and light-front helicity $\Lambda$ reads
\begin{align}
\ket{P,\Lambda} & =
\ket{P,\Lambda}_{3q} + \ket{P,\Lambda}_{3q+g},\label{proton-state}
\end{align}
with
\begin{align}
 &\ket{P,\Lambda}_{3q}  = \sum_{\{\lambda_i\}} \sum_{\{q_i\}}    \int [Dx]_3 \Psi^{\Lambda}_{3q} (\beta,r) 
\, \varepsilon^{c_1c_2c_3}\prod_{i=1}^3\ket{\lambda_i,q_i,c_i,\tilde{k}_i}, \label{LFWF3q}\\
 &\ket{P, \Lambda}_{3q+g} = \sum_{\{\lambda_i\}_{i=1}^4} \sum_{\{q_i\}} \int [Dx]_4 \Psi^{\Lambda}_{3q+g} (\beta,r)\, \varepsilon^{dc_2c_3}  T^a_{d,c_1}\ta \prod_{i=1}^3\ket{\lambda_i,q_i,c_i,\tilde{k}_i}\tc \ket{\lambda_4,g,a,k_i}.\label{LFWF3qg}
\end{align}
In Eqs.~\eqref{LFWF3q} and \eqref{LFWF3qg}, $\Psi^{\Lambda}_{3q}$ and $\Psi^{\Lambda}_{3q+g}$  are, respectively, the  LFWF for the $N=3$ and $N=4$ parton Fock state
$\prod_{i=1}^N\ket{\lambda_i,q_i,c_i,k_i}$, with $\lambda_i$ the parton light-front helicity, $q_i=u,d$ and $g$ the quark and gluon flavor index, $c_i$ the parton color index, and $k_i$ the parton momentum.
For the argument of the LFWFs, we used a collective notation, with  $\beta=(\{\lambda_i\};\{q_i\})$ and  $r=\{\tilde k_i\}$, where 
 $\tilde k_i=(k^+_i=x_i P^+,\bm{k}_{\perp,i})$. 
 Furthermore, the sum over the color indexes is understood, and, then, using also  the sum over the flavor index,  the color matrix in  
 Eq. (\ref{LFWF3qg}) can be saturated with the color index of the first quark only. 
 
The integration measures are defined as:
\begin{align}
[D\tilde k]_{N}& = \frac{[dx]_{N}[d\vetp{k}]_{N}}{\prod_{i=1}^N\sqrt{x_i}},\\
[dx]_{N}& = \delta\ta 1-\sum_{i=1}^N x_i\tc \prod_{i=1}^N dx_i,\\
[d\vetp{k}]_{N} &= \ta \frac{1}{2(2\pi)^3}\tc^{N-1}\delta^2\ta \sum_{i=1}^N \bm{k}_{\perp,i}\tc  \prod_{i=1}^N d\bm{k}_{\perp,i}.
\end{align}
The partial contribution of the $N$-parton Fock state is defined as
\begin{align}
_{N}\langle P'\Lambda'|P\Lambda\rangle_{N}=2P^+(2\pi)^3\delta(P^+-P'^+)\delta^2(\bold{P}'_\perp-\bold{P}_\perp)\delta_{\Lambda,\Lambda'}\, P_{N},
\end{align}
where $P_N$ is the probability to find the $N$-parton state in the proton.

Using the expressions of Eqs.~\eqref{proton-state}-\eqref{LFWF3q}  for the proton state in the definition (\ref{Etilde:def}),   the LFWF overlap representation of  $\tilde e^q$ reads
\begin{align}
 \tilde e^q(x,k_\perp)&= -\frac{\sqrt{2}g_s}{Mx} \int \frac{[d\tilde k]_4}{\sqrt{x_4}}  T^a_{d,c_1}\varepsilon^{dc_2c_3} \varepsilon^{c'_1c'_2c'_3}
 \sum_\Lambda \sum_{ \{\lambda'_i\} } \sum_{ \{q'_i\} } \sum_{ \{\lambda_i\} } \sum_{ \{q_i\} } \sum_{j,j'=1}^3 \delta_{q'_{j'} q}\,\delta_{q_jq}\,\delta_{\lambda'_{j'},-\lambda_j}
\,\delta_{\lambda_4,2\lambda'_{j'}} \notag \\
& \times T^a_{c'_{j'}c_j}   \left( \prod_{\substack{ i=1\\ i\neq j}}^3 \prod_{\substack{ i'=1\\ i'\neq j'}}^3\delta_{c'_{i'}c_i} \,\delta_{q'_{i'}q_i}\,\delta_{\lambda'_{i'}\lambda_i}\right)
  \delta(x-x_j-x_4)\,\delta^2\ta\bm{k}_\perp - \bm{k}_{\perp,j} - \bm{k}_{\perp,4} \tc \notag \\
& \times \Psi_{3q}^{\Lambda *} \Bigg( \{\lambda'_i\}; \{q'_i\}; \big( \tilde{k}'_{j'} = \tilde{k}_j+\tilde{k}_4\big) ; \{\tilde{k}'_{i'} = \tilde{k}_i\}_{i\neq j}^{i'\neq j'} \Bigg) 
\Psi_{3q+g}^{\Lambda} \Bigg( \{\lambda_i \}; \{q_i \}; \{\tilde{k}_i\}\Bigg) \ .\label{overlap-etilde}
\end{align}
 Eq.~\eqref{overlap-etilde} involves the overlap of LFWFs for the $3q$ and the $3q+g$ Fock states, giving direct information on the quark-gluon correlations inside  the proton.
The momentum of the active quark in the $3q$ LFWF
 is set to the external momentum $(k^+=xP^+,\vetp{k})$ that is also the sum of the gluon momentum and one of the quark momentum in the $3q+g$ LFWF. 
 This makes evident  the partonic interpretation of the $\tilde e$ term as an 
 interference between scattering from a coherent quark-gluon pair and from
a single quark~\cite{Efremov:2002qh,Jaffe:1991ra,Jaffe:1991kp}.

The corresponding results for the LFWF overlap representation of the twist-two contribution to $e^q$, related to the unpolarized distribution $f^q_1$,  are given in terms of the sum of the  squares
 of the  $N$-parton LFWFs, according to the density interpretation of the twist-two distributions (see, e.g., Ref.~\cite{Pasquini:2008ax}).

\subsection{Relation to nucleon distribution amplitudes}

In this section, we present a parametrization of the proton LFWFs in terms of leading-twist (twist-three) and next-to-leading-twist (twist-four) proton DAs.
To this aim, we consider the component of the proton state corresponding to vanishing total orbital angular momentum of the partons~\cite{Braun:2011aw,Ji:2003yj,Ji:2002xn}, 
i.e.\footnote{
In Ref. \cite{Ji:2003yj}  one finds two LFWFAs for the $3q$ component with $L_z=0$. However, one of them is suppressed, being related to a next-to-next to leading order DA, and will be neglected in the following.} 
\begin{align}
\ket{P,+}_{3q}^{L_z  = 0} & = \frac{-\varepsilon^{ijk}}{\sqrt{6}}\int [Dx]_{123} \Psi^{(0)}(1,2,3)  \Big( u^{\dagger}_{\uparrow , i}(1)u^{\dagger}_{\downarrow , j}(2)d^{\dagger}_{\uparrow , k}(3)  -u^{\dagger}_{\uparrow , i}(1)d^{\dagger}_{\downarrow , i}(2)u^{\dagger}_{\uparrow , k}(3)\Big) \ket{0},\notag\\
&\label{3qstate} \\
\ket{P,+}_{3q+g\downarrow}^{L_z=0}   &= \varepsilon^{ijk}\int [Dx]_{1234}\Psi^{\downarrow}(1,2,3,4)\, T^{a}_{\sigma i}g^{a\dagger}_{\downarrow}(4) u^{\dagger}_{\uparrow,\sigma}(1)u^{\dagger}_{\uparrow,j}(2)d^{\dagger}_{\uparrow,k}(3)\ket{0}, \notag\\
&\label{3qgdown}\\
\ket{P,+}_{3q+g\uparrow}^{L_z=0}  &= \varepsilon^{ijk}\int [Dx]_{1234}  \Bigg[ \Psi^{1\uparrow}(1,2,3,4)\,T^{a}_{\sigma i} g^{a\dagger}_{\uparrow}(4) u^{\dagger}_{\downarrow,\sigma}(1) \Big(u^{\dagger}_{\uparrow,j}(2)d^{\dagger}_{\downarrow,k}(3) 
 - d^{\dagger}_{\uparrow,j}(2)u^{\dagger}_{\downarrow,k}(3)\Big)  \notag\\
 &+ \Psi^{2\uparrow}(1,2,3,4) \, T^{a}_{\sigma j} g^{a\dagger}_{\uparrow}(4)u^{\dagger}_{\downarrow,i}(1)\Big(u^{\dagger}_{\downarrow,\sigma}(2)d^{\dagger}_{\uparrow,k}(3)  - d^{\dagger}_{\downarrow,\sigma}(2)u^{\dagger}_{\uparrow,k}(3)\Big)\Bigg]\ket{0}, \notag\\
&\label{3qgup}
\end{align}
where $g_\uparrow$ ($g_\downarrow$) denotes the gluon state with positive (negative) light-front helicity.

With respect to the expansion in terms of LFWFs of Eqs.~\eqref{LFWF3q} and \eqref{LFWF3qg}, here
 the flavour and helicity structure of the parton composition is made explicit, and the light-front wave amplitudes (LFWAs) $\psi^{(j)}$ are scalar functions, which depend only on the parton momenta, i.e. the argument $i=1,2,3,4$ stands for $\tilde k_i$. The dependence of the LFWAs  on the factorization scale is implicit. 
The proton state with negative helicity is obtained from Eqs.~\eqref{3qstate}-\eqref{3qgup} by applying the light-front parity transformation $Y$, which corresponds to a parity operation followed by a 180$^{\mathrm{o}}$ rotation around
the $y$ axis. It acts on a state of momentum $P$ and light-front helicity $\Lambda$ as
\begin{equation}
Y|P\Lambda\rangle=(-1)^{s-\Lambda}\eta|P-\Lambda\rangle,\label{lf-parity}
\end{equation}
where $s$ is the total spin of the state and $\eta$ is the intrinsic parity of the hadron ($\eta=+1$ for a proton and a quark state, $\eta=-1$ for a gluon state).

The  representation of $\tilde{e}^q(x,k_\perp)$ in terms of overlap of the LFWAs in Eqs.~\eqref{3qstate}-\eqref{3qgup} is reported in App.~\ref{appendix-2}.

Following Ref.~\cite{Braun:2011aw}, we can write the LFWAs using the following factorized form
\begin{align}
\Psi^{(0)}(1,2,3)  &= \frac{1}{4\sqrt{6}}\phi(x_1,x_2,x_3)\Omega_3(a_3,x_i , \bm{k}_{\perp,i}), \label{lfwa-1}\\
\Psi^{\downarrow}(1,2,3,4)  
& = \frac{1}{\sqrt{2x_4}}\phi_g(x_1,x_2,x_3,x_4)\Omega_4(a_\downarrow,x_i , \bm{k}_{\perp,i}), \\
\Psi^{1\uparrow}(1,2,3,4)  
&  = \frac{1}{\sqrt{2x_4}}\psi^1(x_1,x_2,x_3,x_4)\Omega_4(a_\uparrow,x_i , \bm{k}_{\perp,i}), \\
\Psi^{2\uparrow}(1,2,3,4)   
&  = \frac{1}{\sqrt{2x_4}}\psi^2(x_1,x_2,x_3,x_4)\Omega_4(a_\uparrow,x_i , \bm{k}_{\perp,i}), \label{lfwa-4}
\end{align}
where the $\vetp{k}$-dependence of the functions $\Omega_N$ is assumed to be Gaussian, i.e.
\begin{align}
&\Omega_N(a_N,x_i,\bm{k}_{\perp,i}) = \frac{\ta 16\pi^2a^{2}_{N}\tc^{N-1}}{\prod_{i=1}^Nx_i} \exp\left[-a^{2}_{N}\sum_{i=1}^N \frac{k_{\perp,i}^2}{x_i}\right],\label{omega}
\end{align}
with
 the normalization:
\begin{equation}
\int [d\vetp{k}]_{1...N}\Omega_N = 1.
\end{equation}
The pure $x_i$-dependent part of the LFWAs in Eqs.~\eqref{lfwa-1}-\eqref{lfwa-4} can be directly related to the proton  DAs~\cite{Braun:2011aw,Diehl:1998kh,Bolz:1996sw}.
In particular,  the LFWA which enters  the $3q$ component of the proton state coincides with the leading twist-three proton DA, i.e.
 \begin{align}
 &  \phi(x_1,x_2,x_3) = \Phi_3(x_1,x_2,x_3).
 \end{align}
 The LFWAs for $3q+g$ Fock-state are related to the next-to-leading-oder twist-four DAs~\cite{Braun:2008ia} by\footnote{Eq.~\eqref{phig} differs by a sign from the 
 corresponding relation in Ref.~\cite{Braun:2011aw}. The opposite sign comes from using correctly the light-front parity transformation~\eqref{lf-parity} to flip the light-front helicity of the partons, according to the definition of the $\Xi_4^g$ DA.}
  \begin{align}
\phi_g(x_1,x_2,x_3,x_4)  &= \frac{M}{96g_s} \ta 2\Xi_4^g(x_1,x_2,x_3,x_4) + \Xi_4^g(x_2,x_1,x_3,x_4)\tc ,\label{phig}\\
\psi^1(x_1,x_2,x_3,x_4)   & = -\frac{M}{96g_s} \ta 2\Psi_4^g(x_2,x_1,x_3,x_4) + \Phi_4^g(x_1,x_2,x_3,x_4)\tc, \\
\psi^2(x_1,x_2,x_3,x_4)  &= \frac{M}{96g_s}  \ta 2\Phi_4^g(x_1,x_2,x_3,x_4) + \Psi_4^g(x_2,x_1,x_3,x_4)\tc .
\end{align}
The distribution amplitudes can be expanded into a basis of orthogonal polynomials, leaving all the factorization-scale dependence in the coefficients of the expansion. 
For the three-quark DA, we keep the first few terms of the conformal expansion, corresponding to~\cite{Braun:2008ia}
\begin{align}
 \Phi_3(x_1,x_2,x_3) =  120f_N(\mu)x_1x_2x_3 \ta 1 + A(\mu)(x_1-x_3) + B(\mu)(x_1 + x_3 - 2x_2)\tc .\label{3q-da}
\end{align}
For the twist-four DAs, we adopt  the asymptotic form, given by
\begin{align}
\Xi_4^g(x_1,x_2,x_3,x_4)& = \frac{8!}{6}x_1x_2x_3x_4^2\lambda_1^g(\mu), \label{3qg-da1}\\ 
\Psi_4^g(x_1,x_2,x_3,x_4)
&= \frac{8!}{4}x_1x_2x_3x_4^2\ta \lambda_2^g(\mu) +\frac{\lambda_3^g(\mu)}{3}\tc ,\label{3qg-da2}\\ 
\Phi_4^g(x_1,x_2,x_3,x_4) 
& = -\frac{8!}{4}x_1x_2x_3x_4^2\ta \lambda_2^g(\mu) -\frac{\lambda_3^g(\mu)}{3}\tc .\label{3qg-da3}
\end{align}
The normalization constant $f_N(\mu)$ in Eq.~\eqref{3q-da} and the  parameters $\lambda_i^q(\mu)$ of the expansion in Eqs.~\eqref{3qg-da1}-\eqref{3qg-da3} have been evaluated
using  QCD sum rules \cite{Braun:2011aw} at the scale $\mu=1$ GeV and in the chiral limit of vanishing quark masses, while   the ``shape" parameters  $A,B$  in Eq.~\eqref{3q-da}  have been calculated  on the lattice from the DA moments~\cite{Braun:2008ur,Braun:2010hy}.  

In our approach, we introduce an explicit dependence on the  mass of the ``constituent" partons. This amounts to  effectively taking into account the non-perturbative nature of the proton state modelled in terms of few Fock-state components.
Accordingly, we use  
 the Brodsky-Huang-Lepage prescription~\cite{BHL}, and modify the $\bold{k_\perp}$-dependent part of the LFWA 
 by replacing $k_\perp^2$ with $k_\perp^2 + m^2$, i.e.,  we  use
 \begin{align}
\Omega_N(a_N,x_i,\bm{k}_{\perp,i})  = \frac{\ta 16\pi^2a^{2}_{N}\tc^{N-1}}{\prod_{i=1}^Nx_i} \exp\left[-a^{2}_{N}\sum_{i=1}^N \frac{k_{\perp,i}^2+m_i^2}{x_i}\right].\label{omega-mass}
\end{align}
For the $x$-dependent part of the LFWAs, we keep the same expressions in terms of proton DAs as in Eqs~\eqref{3q-da}-\eqref{3qg-da3}. Since we moved away from the chiral limit, we do not use 
the results of QCD sum rule and lattice QCD for the coefficients of the DAs.
Instead, we treat them as free parameters, as specified in the following.

\section{Results}
\label{sec:3}
\subsection{Fixing the parameters from twist-two parton distributions}
\label{sub-sec:3.1}
The parameters of the LFWFs are fixed to reproduce available phenomenological parametrizations for the unpolarized PDF $f_1$.
In particular, we used the results of the MMHT2014 parametrization~\cite{Harland-Lang:2014zoa} at $Q^2=1$ GeV$^2$, i.e.  the same scale 
of the results from lattice and QCD sum rules.
The valence-quark contribution to $f_1$ is fitted in  the range $x\in(0,1)$, while we limit to $x\in(0.4,1)$  for the fit of the  gluon unpolarized PDF. 
In the last case, we did not include lower values of $x$, because in that region our nonperturbative constituent parton model is not able to catch the low-$x$ dynamics of the gluons 
and, at the same time, the phenomenological parametrizations are not so well constrained  at low $x$ and low scales.

In the fit procedure, the renormalization constant $f_N$ of the leading-twist DA in Eq.~\eqref{3q-da} and the parameters $\lambda_1^g$, $\lambda_2^g$ and $\lambda_3^g$ of the 
next-to-leading twist DAs in Eqs.~\eqref{3qg-da1}-\eqref{3qg-da3} are considered to be distributed as Gaussian variables with means and standard deviations given by the  estimates of the QCD sum rules and lattice calculations.
The shape parameters $A$ and $B$  in Eq.~\eqref{3q-da} are sampled as uniform variables within a larger range than the error band of  the lattice calculations. This allows us to having more flexibility to take into account   the effects of finite quark and gluon masses. 
The  parton masses along with  the width of the $\bold{k_\perp}$ distributions enter  only in the parametrization of the functions $\Omega_N$  in Eq.~\eqref{omega-mass}.
The parton masses are fitted with  the constraint 
\begin{equation}
0\leq 3m_q + m_g \leq M.
\end{equation}
Furthermore, we  assume the same width for the $k_\perp$-distributions  of the gluon with positive and negative helicity, i.e.
\begin{equation}
a_\downarrow = a_\uparrow = a_4,
\end{equation}
and we take  $a_4$ along with the width $a_3$ for the $3q$ state as additional free parameters. 
The  fitted values of the coefficients of the leading-twist and next-to-leading-twist DAs are shown
in Table~\ref{Tab1}, in comparison with the values obtained from QCD sum rules in the chiral limit and lattice calculations.
For the quark and gluon mass we obtain
\begin{align}
m_q = 0.161\text{ GeV}, \qquad
m_g = 0.050\text{ GeV}.\label{mass-fit}
\end{align}
For the widths,  the fit results are
\begin{align}
a_3 = 0.85 \text{ GeV}^{-1}, \qquad
a_4 = 0.92 a_3, \label{width-fit}
\end{align}
which correspond to have a quark-gluon state slightly
more compact in the transverse-momentum space as compared to the
valence three-quark configuration.

We note that the two sets of parameters in Table~\ref{Tab1} are consistent, within the uncertainty bands, except for the coefficients $A$ and $B$. However, we found very small sensitivity of the fitted PDFs to the $A$ and $B$ parameters.

\begin{table*}[b!]
\centering
\begin{tabular}{c|ccc|ccc}
\hline
 & \multicolumn{3}{c|}{$3q$}&\multicolumn{3}{c}{$3q+g$}  \\
 \hline
 &$f_N$ & $A$ & $B$  &$\lambda_1^g$  & $\lambda_2^g$  & $\lambda_3^g $ \\
 & (10$^{-3}$ GeV$^2$)& &  &(10$^{-3}$ GeV$^2$) &(10$^{-3}$ GeV$^2$) &(10$^{-3}$ GeV$^2$) \\
 \hline
fit& 4.68 & $1.14$  & $0.50 $ & $2.79$ & $1.33$  & $ 0.36$ \\
 \hline
lattice QCD& \multirow{2}{*}{$5.0\pm0.5$} & \multirow{2}{*}{$[0.85,0.95]$}  & \multirow{2}{*}{$[0.23,0.33]$} & \multirow{2}{*}{$2.6\pm 1.2$} & \multirow{2}{*}{$2.3\pm 0.7$}  & 
\multirow{2}{*}{$0.54 \pm 0.2$}\\
and QCD sum rules &&&&&&\\
 \hline
\end{tabular}
\caption{Comparison between the fit results of this work and the lattice results~\cite{Braun:2008ur,Braun:2010hy} for the coefficients of the twist-three DA in Eq.~\eqref{3q-da},
parametrizing the $3q$ LFWAs, and the 
QCD sum rules estimates~\cite{Braun:2011aw} of
the coefficients of the twist-four DAs  in Eqs.~\eqref{3qg-da1}-\eqref{3qg-da3}, parametrizing the  $3q+g$ LFWAs.
\label{Tab1}}
\end{table*} 

\begin{figure*}[t!]
\hspace{-0.6 truecm}
\includegraphics[width = 0.35\textwidth]{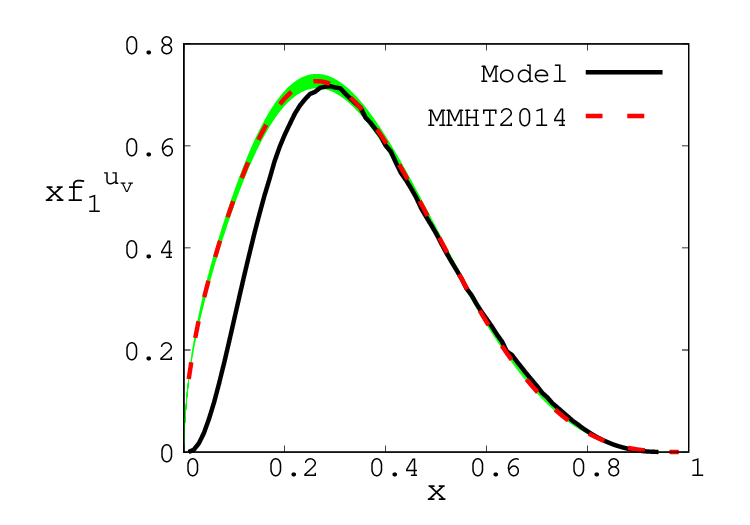}
\hspace{-0.6 truecm}
\includegraphics[width = 0.35\textwidth]{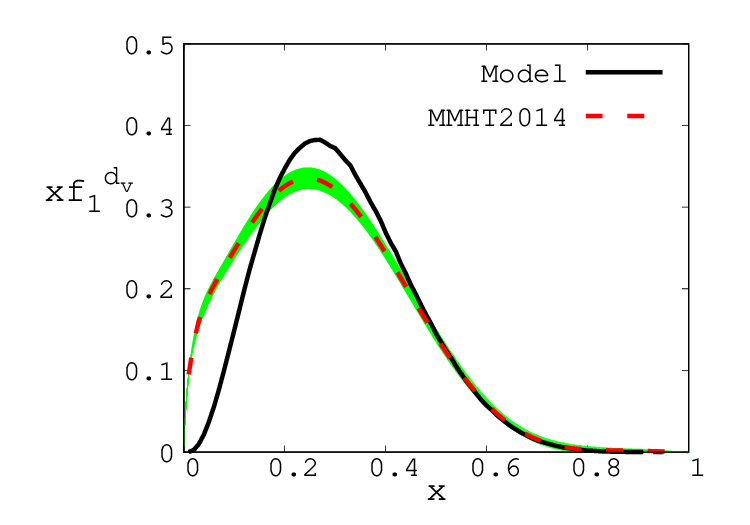}
\hspace{-0.6 truecm}
\includegraphics[width = 0.35\textwidth]{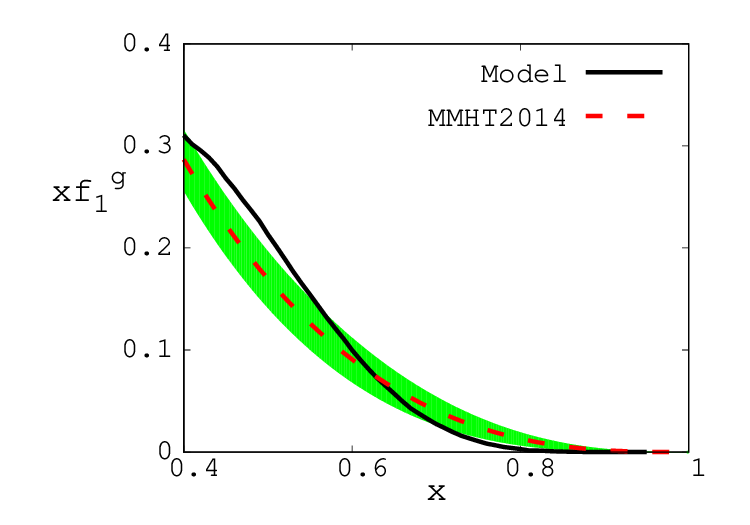}
\caption{The unpolarized PDF $x f_1(x)$ for the valence up-quark (left panel), for the valence down-quark  (middle panel) and  for the gluon (right panel) as function of $x$ at $Q^2=1$ GeV$^2$. Solid curve: results from the light-front model of this work, with the fit parameters in Table~\ref{Tab1} and in Eqs.~\eqref{mass-fit} and \eqref{width-fit}. Dashed curve: results from the parametrization of Ref.~\cite{Harland-Lang:2014zoa}, with the corresponding uncertainty bands.}
\label{fig1}
\end{figure*}

The fit results for the unpolarized parton distributions for the valence up and down quarks,  and for  the gluon are shown
in Fig.~\ref{fig1}, in comparison with the MMHT2014 phenomenological parametrizations~\cite{Harland-Lang:2014zoa} at  $Q^2=1$ GeV$^2$.
The results for the quark PDFs reproduce very well  the MMHT2014 parametrization at larger $x$ (from $x\ge 0.2$, in the case of the up quark, and $x\ge 0.4$, for the down quark), while 
they  have a much faster fall-off for $x\rightarrow 0$ than the Regge-motivated behavior of the parametrization.
The results for the gluon PDF, in the  fitted range of $x\ge 0.4$ reproduce well  the MMHT2014 parametrization.
Furthermore, for the probability of the $3q$ and $3q+g$ components of the proton state we find the following results 
\begin{align}
P_{3q}&=0.18,\\
P_{3q+g}&=P_{3q+g_\uparrow}+P_{3q+g_\downarrow}=0.38,
\end{align}
which are  consistent with the values of Ref.~\cite{Braun:2011aw}.

\subsection{Twist-three distributions}

\begin{figure*}[t]
\hspace{-0.6 truecm}
\includegraphics[width = 0.5\textwidth]{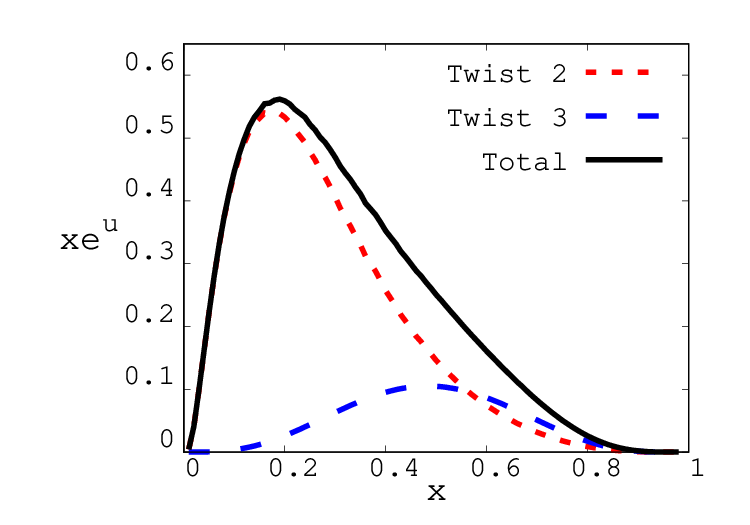}
\hspace{-0.6 truecm}
\includegraphics[width = 0.5\textwidth]{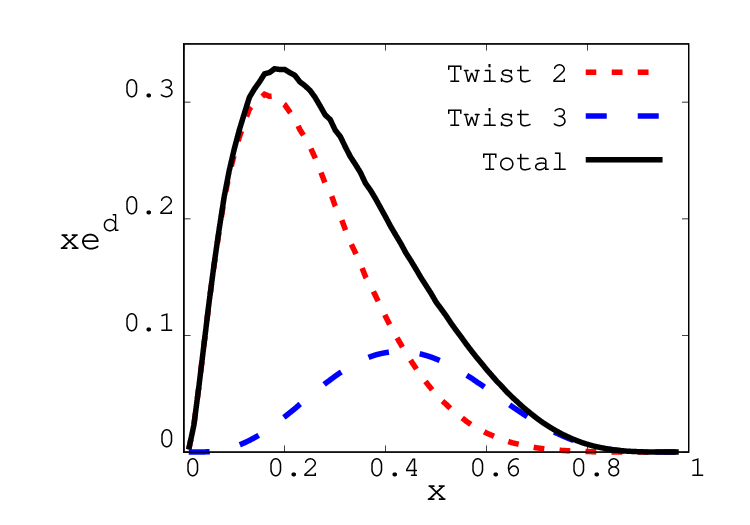}
\caption{Results for the PDF $xe^q(x)$ as function of $x$ for the up  (left panel) and down (right panel) quark. Red short-dashed curve: twist-two contribution; blue long-dashed curve:  pure twist-three contribution; solid curve: total results, sum of the twist-two and twist-three contributions. }
\label{fig:up-down}
\end{figure*}

Using the explicit parametrization for the LFWAs given in Sec.~\ref{sub-sec:3.1}, and the LFWF overlap representation in App.~\ref{appendix-2}, we obtain the results shown in Figs.~\ref{fig:up-down} for the PDF $xe^q(x)$ of up and down quarks. 
The red short-dashed curves show the contribution from the twist-two term, while the blue long-dashed curves correspond to the genuine twist-three contributions.
The relative size of twist-two and genuine twist-three contributions depends on the  quark-mass parameter,  which enters  as proportionality constant that weights the twist-two term in Eq.~\eqref{genDecomp-PDF}.
In our model calculation, the partons' mass also enter 
the functions $\Omega_N$  in Eq.~\eqref{omega-mass}. 
However, in this case,  the  dependence on the partons' mass is such that it does not affect
the relative size of the the twist-two and twist-three contributions to $xe^q(x)$, and slightly changes their behaviour as function of $x$.

We also note that the twist-two and genuine twist-three terms have a quite different $x$-dependence.
The twist-two contribution is peaked at $x\approx 0.2$, with a fast fall-off at larger $x$, more pronounced in the case of up quarks than down quarks.
On the other side, the $x\tilde e^q(x)$ contribution is peaked at $x\approx 0.5$,  and becomes the dominant contribution at larger $x$, especially for down quarks.
We also note that the pure twist-3 contributions for the up and down quarks have very similar size, whereas the twist-2 contribution for the up quark is approximately twice as large as the twist-2 contribution for the down quark (note the different scales on the vertical axis).

\begin{figure}[t!]
\includegraphics[width = 0.5\textwidth]{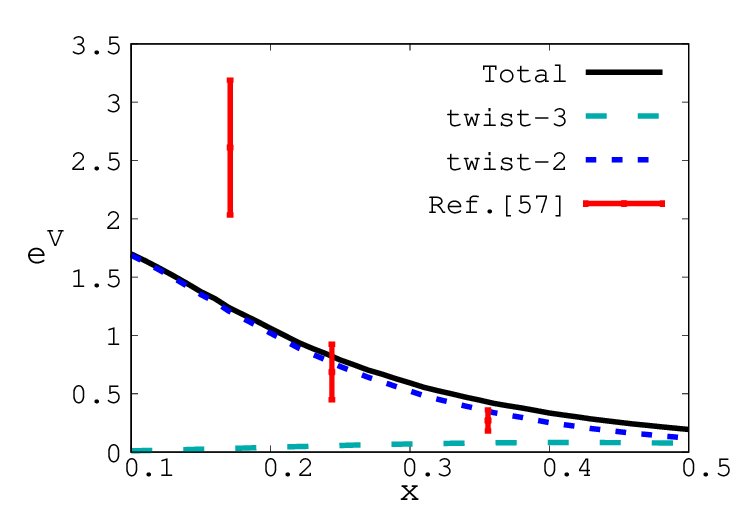}
\caption{Predictions for the combination $e^V=\frac{4}{9} e^u(x) - \frac{1}{9}e^d(x)$ at the scale $Q^2 = 1$ GeV$^2$, in comparison with the extraction of Ref. \cite{Courtoy:2014ixa} at the scale $Q^2 = 1.5$ GeV$^2$. The dashed curve shows the contribution of the twist-two term of $e$, and the solid curve corresponds to the total contribution from the sum of the twist-two and twist-three terms of $e$.}
\label{fig:compl}
\end{figure}

Recently, the CLAS collaboration has reported preliminary results of a measurement of the beam asymmetry in di-hadron SIDIS, using
a longitudinally polarized 6 GeV electron beam off an unpolarized proton target~\cite{Courtoy:2014ixa,E12-06-112B}. 
These data have been analyzed to extract the following flavor combination of the valence-quark contributions to $e^q(x)$:
\begin{equation}
e^V(x) = \frac{4}{9} e^{u_v}(x) - \frac{1}{9}e^{d_v}(x).
\end{equation}
In Fig.~\ref{fig:compl}, the preliminary CLAS data points at the scale $Q^2=1.5$ GeV$^2$ are compared with our model predictions at the scale $Q^2=1$ GeV$^2$. The model results are also split in the  contributions from twist-two and genuine twist 3-terms, corresponding to the short-dashed blue curve and the long-dashed cyan curve, respectively.
Our results are in quite good agreement with the experimental extraction at the two higher-$x$ bins, but they are not able to reproduce the observed fast rising at lower $x$.
This could be due to a lack of our model, that, according to the fit results for the unpolarized PDF $f_1$, shown in Fig.~\ref{fig1},   is less reliable in the lower-$x$ region.
We also notice that the genuine twist-three contribution in the considered $x$-range is very small, supporting the results within the light-front constituent-quark picture that was used
in Ref.~\cite{Lorce:2014hxa} and showed to be able to reproduce the results of the CLAS data at  higher $x$.
 However, one should bear in mind that these data are still preliminary and have unestimated systematic uncertainties.


\subsection{Transverse momentum}

\begin{figure*}[!t]
\hspace{-0.6 truecm}
\includegraphics[width = 0.5\textwidth]{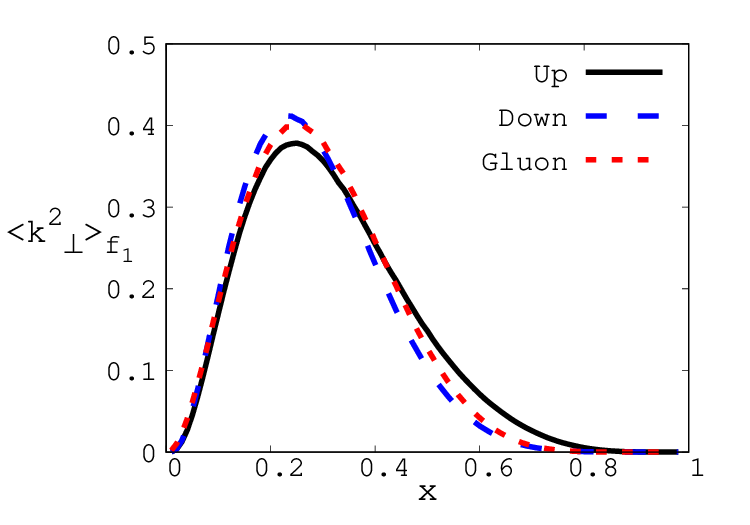}
\hspace{-0.6 truecm}
\includegraphics[width = 0.5\textwidth]{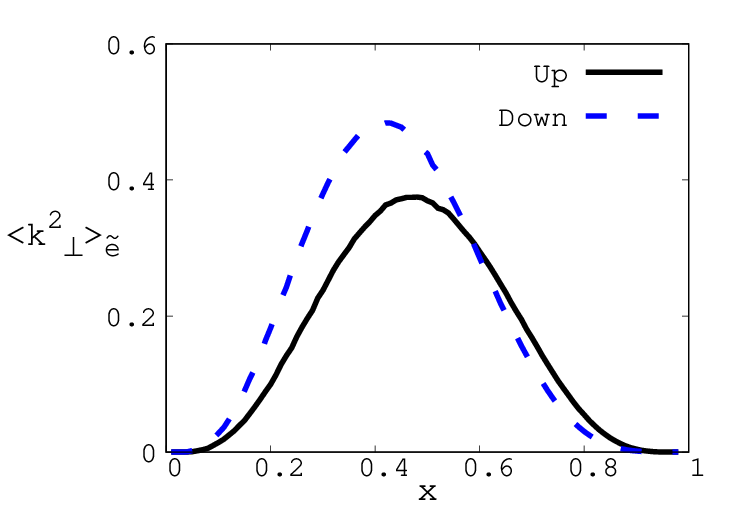}
\caption{Results for the $x$-dependence of the mean squared transverse momentum in Eq.~\eqref{Kperp2Def}  of the unpolarized TMD $f_1$ (right panel) and the TMD $\tilde e$ (left panel) at the model scale of $Q^2=1$ GeV$^2$.
The solid black curves and the long-dashed blue  curves correspond to the up and down quark contributions, respectively, while the long-dashed red curve in the case of $f_1$ shows 
the gluon contribution.}
\label{kperp_x}
\end{figure*}

Next, we turn our attention to the $k_\perp$ dependence of the TMDs.
We define the $x$-dependent mean squared transverse momentum of a generic TMD $j\ta x, k_\perp\tc$ as follows
\begin{equation}
\braket{k_\perp^2}_j(x) = \frac{\int d\vetp{k} \ k_\perp^2 j\ta x, k_\perp\tc}{\int dx\,d\vetp{k} j\ta x, k_\perp\tc}.
\label{Kperp2Def}
\end{equation}
The corresponding results for the quark and gluon  unpolarized distribution $f_1\ta x, k_\perp\tc$ 
and for the quark TMD $\tilde e\ta x, k_\perp\tc$ are shown in the left and right panel of Fig.~\ref{kperp_x}, respectively.
The quark results refer to the valence-quark contribution, since in our model we neglect the sea quarks.
The results for the gluon are obtained from the $3q+g$ component of the proton LFWF, and therefore correspond to the  intrinsic non-perturbative gluon contribution.
All the results refer to the model scale of $Q^2=1$ GeV$^2$.
We find that in the case of 
 $f_1$, the size and the $x$-dependence is very similar for  up and down quarks, and for gluons. 
They are all peaked at $x\approx 0.3$, and fall down rapidly at higher $x$, with a very similar slope  in the case of the down quarks and the gluon.
The similar behavior for down quark and gluons is an indication that the results for the down quark contribution to the $f_1$ TMD are more dominated by the $3q+g$ component of the LFWF than in the case of up quark.
For the quark contribution, there exists a recent 
extraction, based on a fit of the unpolarized TMD $f_1(x,k_\perp)$ to   available experimental data  measured in SIDIS, Drell-Yan and $Z$ boson  production~\cite{Bacchetta:2017gcc}. This extraction assumes no quark-flavour dependence and uses data  in  the range of $5\cdot 10^{-2}\le x\le 0.4$. Therefore, the corresponding results for the mean squared transverse momentum  get a sizeable contribution from sea-quarks, which results in a $x$-dependence quite different from the bell-shaped structure we find in our calculation.
Nevertheless, we find that   our results  are in the same range of values as  the phenomenological extraction.

In the case of the mean squared transverse momenta  of  $\tilde e^q(x,k_\perp)$  we observe a more pronounced quark flavor  dependence 
 w.r.t. to the case of the unpolarized twist-two TMD. Furthermore, they are peaked at higher-$x$ values, with the  contribution from up quarks slightly shifted at larger $x$ w.r.t. to the one from  down quarks.

By integrating  Eq.~\eqref{Kperp2Def} over $x$, we obtain the results for the $\braket{k^2_\perp}$ width of the TMDs.
The results for $f_1$ and $\tilde e$ at the model scale of $Q^2=1$ GeV$^2$ are  presented in Table~\ref{TabPT2}.
They  are very similar for all the partons in the case of $f_1$, while they have an appreciable quark-flavor dependence in the case of $\tilde e$. We also notice  the broader widths for the  $\tilde e$ distribution with respect to $f_1$.
\begin{table}[!h]
\centering
\begin{tabular}{c|ccc|cc}
\hline
 & \multicolumn{3}{c|}{$f_1$}&\multicolumn{2}{|c}{$\tilde e$}\\
 \hline
 &$u$&$d$&$g$ &$u$&$d$  \\
\hline
$\braket{k^2_\perp}$ (GeV$^2$)& $0.138 $  & $0.133 $&  $0.136 $ & $0.158$  & $ 0.196$ \\
 \hline
\end{tabular}
\caption{Results for $\braket{k_\perp^2}$ of the $f_1$ and $\tilde e$ TMDs at $Q^2=1$ GeV$^2$.}
\label{TabPT2}
\end{table} 

\begin{figure}[!t]
\includegraphics[width = 0.47\textwidth]{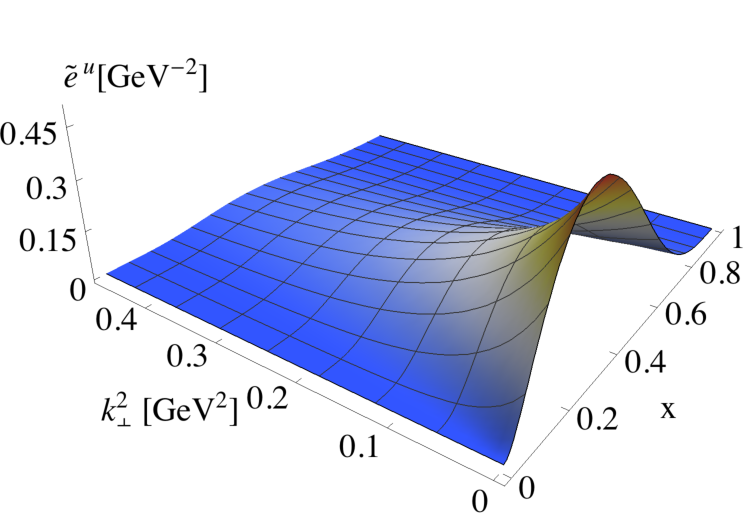}
\hspace{0.5 cm}
\includegraphics[width = 0.47\textwidth]{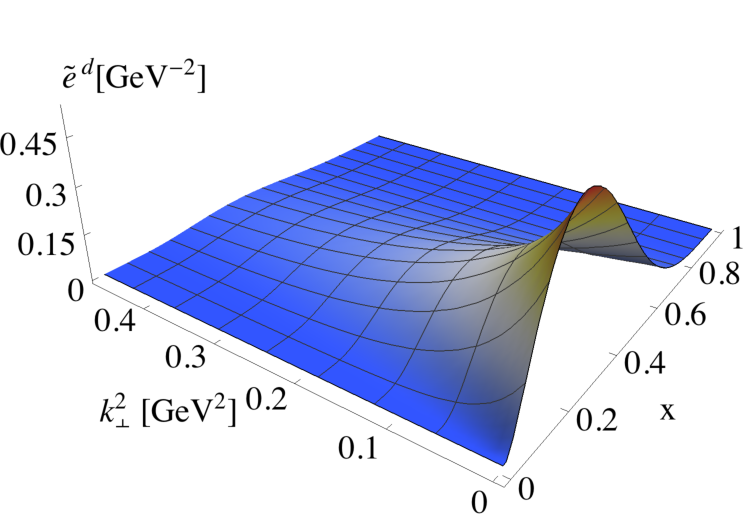}
\caption{$\tilde e$ for the up (left panel) and down (right panel) quark as a function of $x$ and $k_\perp^2$.}
\label{fig:tmd}
\end{figure}

Finally,  in Fig.~\ref{fig:tmd} we show the three-dimensional plots of $\tilde e$
for the up and down quark, as function of $x$ and $k^2_\perp$. For both quark flavors, the $k^2_\perp$-dependence of the distributions  slightly moves away from a Gaussian shape, as expected from our model Ansatz for the $k^2_\perp$-dependence of the LFWFs in Eq.~\eqref{omega-mass}.

\section*{Conclusions}
\label{sec:4}

In this work, we studied the sub-leading structure function $e^q$.

First, we reviewed its model-independent decomposition, which follows from the QCD EOM.
In particular, this decomposition contains a genuine twist-three part, i.e. the ``tilde" term encoding the quark-gluon correlations, a pure twist-two contribution and a singular ($\delta$-like) term.
The singular term is in turn given by a well-know contribution that can be related to the pion-nucleon-sigma term, and an additional term that, instead, is poorly discussed in literature and has been written down  explicitly here for the first time.

Then, we focused on the modeling of the tilde term. To this aim, we constructed a model-independent representation in terms of overlap of LFWFs for the $3q$ and $3q+g$ Fock-components
of the proton state.
For the calculation, we restricted ourselves to the LFWFs corresponding to zero orbital angular momentum of the partons. This allowed us to relate the $3q$ and the $3q+g$ 
LFWFs in the limit of zero-transverse separation to the leading-twist  and next-to-leading-twist DAs of the proton, respectively. 
The transverse-momentum dependence of the LFWFs was built up by assuming a modified Gaussian Ansatz, with explicit dependence
on the mass of the partons.
The DAs and the  transverse-momentum dependent part were parametrized
in such a way to reproduce available phenomenological parametrizations for the unpolarized PDF $f_1$ of quarks and gluons at the scale of $Q^2=1$ GeV$^2$.
The fit gave values for the parametrization of the DAs that are  consistent with lattice calculations~\cite{Braun:2008ur,Braun:2010hy} and predictions from QCD sum rules~\cite{Braun:2011aw}.

With these ingredients, we provided predictions for both the quark PDF $\tilde e^q(x)$ and the quark TMD $\tilde e^q(x,k_\perp)$, in comparison with the corresponding 
twist-two contribution given in terms of $f^q_1$.
In the case of $\tilde e^q(x)$,  we found that the pure twist-three contribution has a distinctive $x$-dependence from the twist-two contribution, and is about 20-30$\%$
of the twist-two contribution  at the peak position, using our model-results for the  quark masses. We also considered
preliminary results from a phenomenological extraction of a particular flavor combination of the valence-quark contribution to $e^q(x)$~\cite{Courtoy:2014ixa}.
 For this quark-flavor combination, our model predictions are almost saturated by the twist-two contribution to $e^q(x)$, and showed 
   a quite good agreement with the extracted results in the large $x$-region.

The results of this work are encouraging for an extension to other subleading-twist TMDs.
However, twist-three TMDs other than $e^q$ involve a transfer of orbital angular momentum between the initial and final proton states.
This requires the modeling of the LFWF components with non-zero orbital angular momentum. Work in this direction is in progress.

\section*{Acknowledgments}

The authors are grateful to A. Bacchetta, V. Braun, C. Lorc\'e, and P. Schweitzer for stimulating discussions, and to T. Liu for helpful advices in using the phenomenological parametrizations of PDFs. 
This work is partially supported by the European Research Council (ERC) under the European Union's Horizon 2020 research and innovation programme (grant agreement No. 647981, 3DSPIN).

\appendix
\section{}
\label{appendix-1}
In this appendix, we show the derivation of the contributions $e^q_m$ in Eq.~\eqref{MassTerm} and $e^q_{\text{tw3}}$ in Eq.~\eqref{GluonTerm}.

We start by considering the matrix elements 
of the operators ${\cal O}_m$ and ${\cal O}_{\text{tw3}}$ in Eqs.~\eqref{Om} and \eqref{Oa}, respectively, which enter the definition~\eqref{Eq:correlator-TMDs2} of the TMD $e^q$:
\begin{align}
&\int \frac{dz^-d\vetp{z}}{(2\pi)^3}e^{iz^-xP^+-i\vetp{z}\cdot\vetp{k}} 
\int_{0^-}^{z^-}d\zeta^- \mathcal{M}_l(\zeta^-,\vetp{z}), \, (l=m,a),\label{matrix-element}
\end{align}
with
\begin{align}
\mathcal{M}_m(\zeta^-,\vetp{z})
&:=-\frac{im}{2} \braket{P|\overline\psi(0)\gamma^+\mathcal{W}_1\ta0^-,\vetp{0};\zeta^-,\vetp{z}\tc\psi(\zeta^-,\vetp{z})|P},\\
\mathcal{M}_{\text{tw3}}(\zeta^-,\vetp{z})&:=\frac{g_s}{2} \Big(
\int_{\infty^-}^{\zeta^-} d\eta^- \braket{P|\overline\psi(0)\mathcal{W}_1\ta0^-,\vetp{0};\eta^-,\vetp{z}\tc G_{\ j}^{+}(\eta^-,\vetp{z})\sigma^{j+}\mathcal{W}_1\ta \eta^-,\vetp{z};\zeta^-,\vetp{z}\tc\psi(\zeta^-,\vetp{z})|P} \notag \\
& +  \int^{\infty^-}_{0^-} d\eta^- \braket{P|\overline\psi(0)\mathcal{W}_1\ta 0^-,\vetp{0};\eta^-,\vetp{0}\tc G_{\ j}^{+}(\eta^-,\vetp{0})\sigma^{j+}\mathcal{W}_1\ta \eta^-,\vetp{0};\zeta^-,\vetp{z}\tc \psi(\zeta^-,\vetp{z})|P}\Big),
\end{align}
where we did not write explicitly $z^+=0^+$ in the argument of the fields.
By integrating  over $\vetp{z}$ and introducing the Fourier-transform  in the variable $\zeta^-$ of the matrix-element $\mathcal{M}_l$, Eq.~\eqref{matrix-element} can be rewritten as
\begin{align}
&\int \frac{dz^-}{2\pi}e^{iz^-xP^+} \int_{0^-}^{z^-}d\zeta^-
\int dp^+e^{-i\zeta^-p^+} \mathcal{M}_l(p^+,\vetp{k}) .\label{matrix-element-2}
\end{align}
The integral over $\zeta^-$ in Eq.~\eqref{matrix-element-2} can be easily performed, giving
\begin{align}
&i\int \frac{dz^-}{2\pi}e^{iz^-xP^+} \int dp^+ \frac{e^{-iz^-p^+} - 1}{p^+}\mathcal{M}_l(p^+,\vetp{k}).\label{matrix-element-3}
\end{align}
Finally, integrating Eq.~\eqref{matrix-element-3} over $z^-$ and changing  the integration variable as $p^+=yP^+$, we obtain
\begin{align}
&\frac{i }{xP^+} \mathcal{M}_l(xP^+,\vetp{k}) - \delta(x)\frac{i}{P^+} \int dy \frac{1}{y}\mathcal{M}_l(yP^+,\vetp{k}).\notag\\
&\label{matrix-element-4}
\end{align}
Eq.~\eqref{matrix-element-4} , for $l=m$ and with  the definition  \eqref{Eq:correlator-TMDs1}  for  $f_1^q$,
corresponds to
the contribution $e_m^q$ in Eq.~\eqref{MassTerm}.
Analogously, Eq.~\eqref{matrix-element-4}, for the matrix element with $l=\text{tw3}$, and with the definition \eqref{Eq:correlator-TMDs2} for  $\tilde e^q$,
 gives the contribution $e^q_{\text{tw3}}$ in Eq.~\eqref{GluonTerm}.
\section{}
\label{appendix-2}
The overlap representation of the $\tilde{e}^q(x,k_\perp)$ TMD in terms of the LFWAs corresponding to the parton configuration with zero partons' orbital angular momentum 
reads
 \begin{align}
  \tilde{e}^u(x,k_{\perp}) &= \frac{4g_s}{Mx\sqrt{3}} \int \frac{[dx]_{1234}[dk]_{1234}}{\sqrt{x_4}}\Bigg\{ - \delta(x-x_4-x_1) \delta^2(\bm{k}_{\perp}-\bm{k}_{\perp 4}-\bm{k}_{\perp 1})  \notag\\
&  \times  \Bigg[  \Bigg( 4\Psi^{(0)*}(-2-3 ,3,2) +2\Psi^{(0)*}(2,3,-2-3) \Bigg) \Psi^{1\uparrow}(1,2,3,4)\notag\\
&+ \Bigg( 2\Psi^{(0)*}(-2-3 ,2,3) + \Psi^{(0)*}(3,2,-2-3)\Bigg)\Psi^{2\uparrow}(1,2,3,4) \notag \\
& - 2 \Psi^{(0)*}(2 , -2-3,3)\Psi^{\downarrow}(1,2,3,4)\Bigg]\notag \\ 
 & +\delta(x-x_4-x_2) \delta^2(\bm{k}_{\perp}-\bm{k}_{\perp 4}-\bm{k}_{\perp 2}) \notag \\
 & \times \Bigg[2\Psi^{(0)*}(-1-3 ,1,3)\Psi^{2\uparrow}(1,2,3,4)  - \Psi^{(0)*}(1 , -1-3 ,3)\Psi^{\downarrow}(1,2,3,4) \Bigg] \notag  \\
 & + \delta(x-x_4-x_3) \delta^2(\bm{k}_{\perp}-\bm{k}_{\perp 4}-\bm{k}_{\perp 3})\,\Psi^{(0)*}(-1-2 ,1,2)\Psi^{1\uparrow}(1,2,3,4) \Bigg\},\label{EtildeUp}
 \end{align}
 for the up quark, and
 \begin{align}
 \tilde{e}^d(x,k_{\perp}) &= -\frac{4g_s}{Mx\sqrt{3}}\int \frac{[dx]_{1234}[dk]_{1234}}{\sqrt{x_4}}\Bigg\{  \delta(x-x_4-x_2) \delta^2(\bm{k}_{\perp}-\bm{k}_{\perp 4}-\bm{k}_{\perp 2})\notag \\
 &\times  2\Psi^{(0)*}(3,1,-1-3)\Psi^{2\uparrow}(1,2,3,4)\notag \\
 & + \delta(x-x_4-x_3) \delta^2(\bm{k}_{\perp}-\bm{k}_{\perp 4}-\bm{k}_{\perp 3})\notag \\
 & \times \Bigg[ \Psi^{(0)*}(2,1,-1-2)\Psi^{1\uparrow}(1,2,3,4)\notag \\
 & - \Bigg( \Psi^{(0)*}(1,-1-2,2) + \Psi^{(0)*}(2,-1-2,1) \Bigg)\Psi^{\downarrow}(1,2,3,4)\Bigg]\Bigg\},\label{EtildeDown}
\end{align}
for the down quark.
By integration of Eqs.~\eqref{EtildeUp} and \eqref{EtildeDown} over $\vetp{k}$, one obtains also the corresponding results for the PDF $e^q(x)$.

\section*{References}

\end{document}